\documentclass[twocolumn,aps,prl,showpacs,amsmath,amssymb,floatfix]{revtex4-1}
\usepackage{graphicx}
\usepackage{dcolumn}
\usepackage{bm,xcolor}
\usepackage{natbib}

\begin{document}

\title{Intermittent ``Turbulence" in a Many-body System}

\author{Guram Gogia}
\author{Wentao Yu}
\author{Justin C. Burton} 

\affiliation{Department of Physics, Emory University, Atlanta, GA, 30322}
\date{\today}

\begin{abstract}
In natural settings, intermittent dynamics are ubiquitous and often arise from a coupling between external driving and spatial heterogeneities. A well-known example is the generation of transient, turbulent ``puffs" of fluid through a pipe with rough walls. Here we show how similar dynamics can emerge in a discrete, crystalline system of particles driven by noise. Polydispersity in particle masses leads to localized vibrational modes that effectuate a transition to a gas-like phase. A minimal model for the evolution of the system’s mechanical energies exhibits quasi-cyclic oscillations, and a single, dimensionless number captures the essential features of the intermittent dynamics, analogous to the Reynolds number for pipe flow.
\end{abstract}

\maketitle

The dynamics of complex systems that are driven away from equilibrium are usually characterized by minor fluctuations around some steady state, which are abruptly punctuated by a ``big jump'' \cite{Goldenfeld1999} leading to a dramatically different state. The climate oscillates between hot and cold regimes over millennia \cite{Petit1999}, lakes switch between high- and low-nutrient regimes on decade-long time-scales \cite{Van2007}, and rain showers pop up for a few minutes during summer afternoons. In spatially extended complex ``ecologies'', such as vegetation \cite{Dodorico2006} and power grids \cite{Nesti2018}, spatio-temporal intermittency emerges due to coupling between external random forcing and the underlying structure. 

A classical physical example of intermittent dynamics is transitional pipe flow \cite{Eckhardt2007}, where laminar fluid flow develops spatially disordered and transient turbulent ``puffs'' for intermediate flow velocities. Since laminar flow in a pipe is linearly stable for all Reynolds numbers  (Re) \cite{Lessen1968}, the emergence of turbulence requires a finite disturbance, such as rough walls \cite{Nikuradze1933} or other localized perturbations \cite{Reynolds1883,Hof2003}. Despite tremendous progress in experiment \cite{Avila2011} and theory \cite{Chate1987,Barkley2016,Shih2016,Manneville2016} to capture the statistical properties of turbulence, the exact mechanism of cooperation between structural perturbations and the fluid flow remains elusive.  The main challenge, as in other complex systems, lies in relating the local spatiotemporal interactions to the system-wide intermittent behavior. 
\begin{figure}[!t]
\begin{center}
\includegraphics[width=3.3 in]{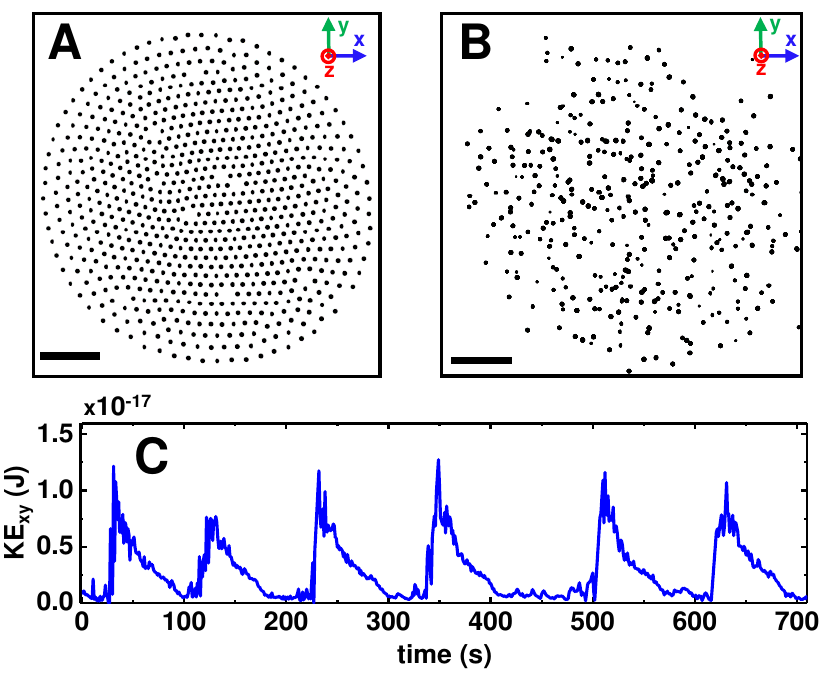}
\caption[]{Intermittent switching occurs over minutes between (A) crystalline and (B) gas-like states. The scale bars correspond to 5 mm. (C) Horizontal kinetic energy per particle in a system comprised of 751 melamine formaldehyde spherical particles (density = 1,510 kg/m$^3$, average radius = 4.73 $\mu$m).
\label{ExpResults} }
\end{center} 
\end{figure}

Here we illustrate this connection in a simple, many-body system that manifests intermittent ``turbulence". A spatially-extended, horizontal layer of charged particles, forced vertically by white noise, continuously switches between crystalline and gas-like (turbulent) states. Particle polydispersity leads to spatially-localized vibrational modes that are necessary for the switching to occur. Upon excitation, these modes nonlinearly couple to other modes and redistribute the energy. Furthermore, we introduce a minimal model for the evolution of the total mechanical energy in the vertical and horizontal directions, reminiscent of stochastic predator-prey systems used to describe turbulence \cite{Shih2016}. In both the simulation and the model, a single dimensionless number that incorporates external driving, dissipation, and disorder, successfully predicts the intermittent dynamical regime.

\begin{figure*}[t!]
\begin{center}
\includegraphics[width=6.7 in]{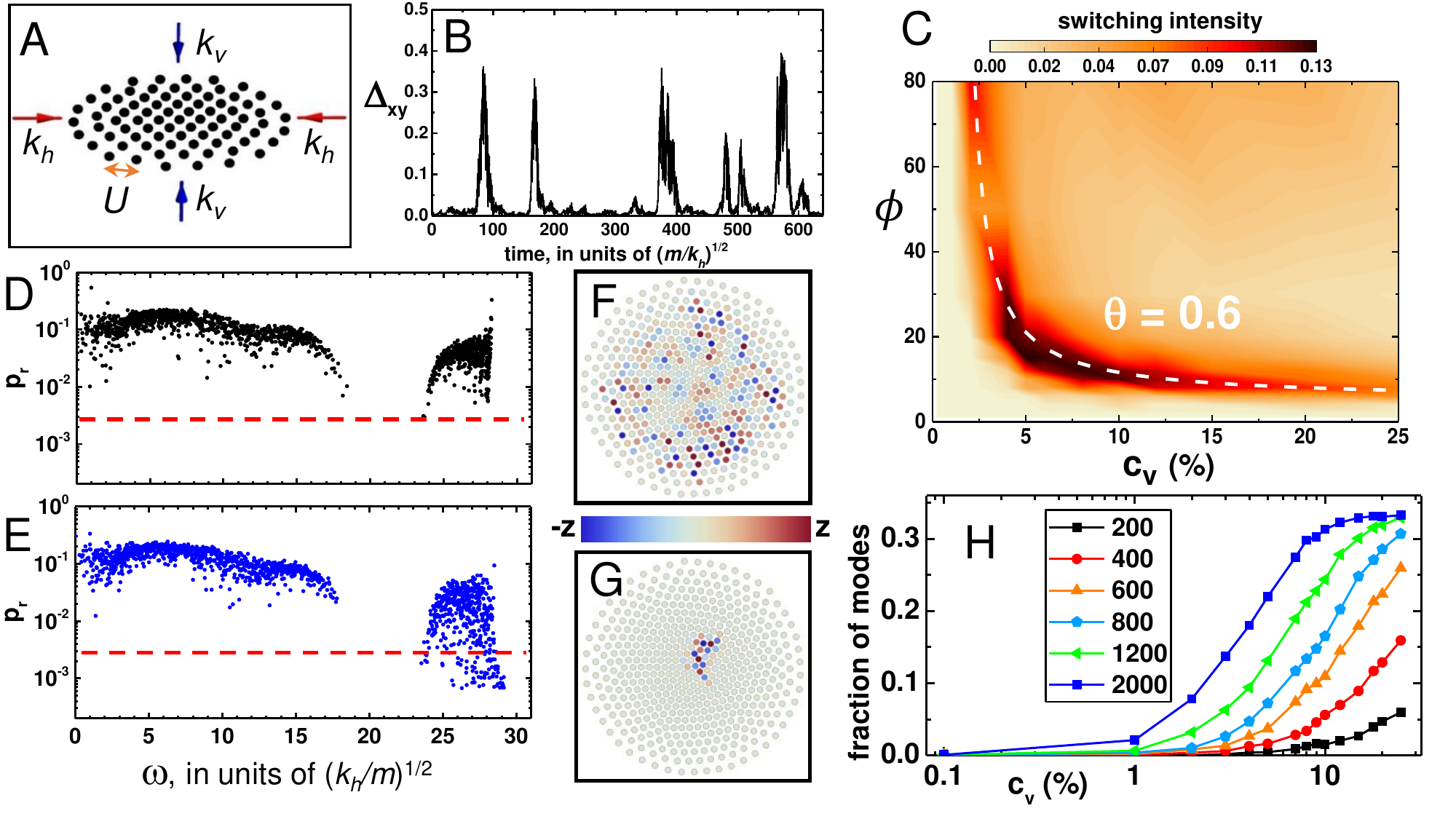}
\caption[]{(A) A single layer of interacting particles is confined horizontally and vertically by harmonic potentials. (B) Temporal evolution of  $\Delta_{xy}$ for a polydisperse system of $N=500$ particles with $c_{v}=4\%$ driven with $\phi=20$. (C) Heatmap showing switching intensity vs. $c_v$ and $\phi$ for a system of 500 particles with $\gamma=0.2$. The dotted line corresponds to $\theta=0.6$ and captures the inverse relation between $\phi$ and $c_v$. Participation ratio ($p_r$) vs. mode frequency for a monodisperse (D) and a polydisperse (E) system. The dashed red line corresponds to the lowest value of $p_r$ in the monodisperse sample. Vertical polarization for two modes (high-$p_r$ (F) and low-$p_r$ (G)) in a polydisperse system with $c_v=4\%$. (H) Fraction of vertical modes with low-$p_r$ as a function of $c_v$ for different values of $k_v/k_h$, as indicated by the legend. 
\label{Sim} }
\end{center} 
\end{figure*}

\textit{Experiments} --- The work described here was primarily motivated by recent experimental observations of emergent, intermittent dynamics in a quasi-2D dusty plasma crystal \cite{Gogia2017}. In the experiment, hundreds of electrostatically-levitated microspheres switch between crystalline (Fig.\ \ref{ExpResults}A) and gas-like states (Fig.\ \ref{ExpResults}B). Fig. \ref{ExpResults}C shows the average kinetic energy per particle in the $xy$ imaging plane. The system is inherently nonequilibrium; energy is sourced from the plasma environment, leading to large-amplitude oscillations of each particle in $z$ \cite{Nunomura1999, Samarian2001, Marmolino2011,Harper2019}. Accompanying numerical simulations indicated that switching could only emerge if the particles have some finite size variation. This quenched disorder lead to localized melting which rapidly spread throughout the crystal. Here we generalize these results and identify the key ingredients for this class of intermittent, ``turbulent" dynamics in a many-body system.

\textit{Numerical Simulations} --- We used a custom molecular dynamics code to simulate $N$ particles that interact via a finite-ranged pair potential, $U(r)=U_{0}\lambda e^{-r/\lambda}/r$, where $U_0$ is the characteristic energy scale, $r$ is the particle separation, and $\lambda$ is the screening length. The particles are spatially confined vertically and horizontally through harmonic potentials, $k_{v}z^2/2$ and $k_{h} (x^2+y^2)/2$, where $k_v$ and $k_h$ are the respective spring constants (Fig.\ \ref{Sim}A). Particles were initially placed at random locations in the $xy$-plane and then quenched to the nearest local potential energy minimum using the FIRE algorithm \cite{Bitzek2006}. The anisotropic confinement ($k_v\gg k_h$) leads to a separation of normal mode frequencies associated with in-plane and out-of-plane motion. In addition, we introduce two non-conservative forces: a hydrodynamic drag force, $\vec{\bf{F}}_d=-\gamma m \vec{\bf v}$, where $\gamma$ is the dissipation rate, and a spatially-uniform Langevin force in the $z$-direction to stimulate vertical oscillations, $\vec{\bf F}_s= \hat{\bf z}w(t)\sqrt{m\phi/\Delta t}$, where $w(t)$ is a Wiener process with zero mean and unit standard deviation, $\phi$ is the power delivered by the noise, and $\Delta t$ is the simulation time step. The particle positions and velocities were advanced in time using velocity-Verlet integration.

For identical particles, the noisy forcing only produced vertical motion in the system's center-of-mass. To realize intermittent dynamics, we introduced quenched disorder by choosing the particle masses from a Gaussian distribution with mean $m$ and coefficient of variation $c_v$. With these parameters, we identify a characteristic mass ($m$), length ($\lambda$), and time ($\sqrt{k_h/m}$) scale in the system. In what follows, all variables have been scaled by these units. The intermittent dynamics manifest as punctuated cascades of energy from the vertical to the horizontal directions, as illustrated by the fractional horizontal kinetic energy, $\Delta_{xy}\equiv KE_{xy}/(KE_{z}+KE_{xy})$, shown in Fig.\ \ref{Sim}B. Crystalline states correspond to $\Delta_{xy}\ll 1$, and gas-like states correspond to $\Delta_{xy}\lesssim 2/3$, where the upper limit corresponds to energy equipartition among the degrees of freedom. However, intermittent dynamics were only observed for a small range of parameters. To illustrate this, we characterized the \textit{switching intensity} in the dynamics by integrating the Fourier transform of $\Delta_{xy}$ over frequencies smaller than 0.02 Hz, which corresponds to intermittency periods longer than 50 s (Fig.\ S3 \cite{supp}). A heatmap of the switching intensity (Fig.\ \ref{Sim}C) indicates a distinct regime of intermittent dynamics characterized by an inverse relationship between $c_v$ and $\phi$. Too much disorder or stochastic forcing leads to a perpetual gas-like state, and too little leads to a stable crystalline state. 

The transition to the gas-like state is rather abrupt since a copious amount of excess energy can be stored in the vertical oscillations (Suppl. Movie \cite{supp}). An analysis of the harmonic vibrational modes in the system illustrates how quenched disorder facilitates this transition \cite{Henkes2012,Bottinelli2017,Burton2016}. For each system, the Hessian matrix was computed about the local potential energy minimum corresponding to the crystalline state (\cite{supp}). The 3$N$ eigenvalues of this matrix correspond to squares of mode frequencies, whereas the normalized eigenvectors are the polarizations of the particle displacements. The localization of each mode was characterized using the participation ratio \cite{Burton2016,supp}. Modes with $p_{r}\lesssim 1$ represent the collective motion of many particles, whereas the modes with $p_{r}\ll1$ correspond to localized modes consisting of a few particles. 

Figure \ref{Sim}D-E shows $p_{r}$ versus angular frequency for a monodisperse (D) and a polydisperse (E) system comprised of $N$ = 500 particles. There are two distinct frequency bands. High frequency modes ($\omega\gtrsim$ 25) correspond to vertical motion, and low frequency modes ($\omega<$ 20) correspond to horizontal motion. A frequency gap separates these bands and suppresses mode coupling. A typical extended vertical mode is shown in Fig.\ \ref{Sim}F. A small amount of quenched disorder in particle mass leads to localized modes (Fig.\ \ref{Sim}G) with $p_r\ll1$ in the vertical frequency band, yet leaves the horizontal frequency band essentially unchanged. Upon excitation with a spatially-uniform (wavevector $k=0$) stochastic force, these low-$p_r$ modes are preferentially excited since they contain low-$k$ Fourier components. These modes are also more susceptible to nonlinear coupling since the relative amplitude of motion between neighboring particles is larger. Consequently, they serve as the progenitors of the energy cascade from vertical to horizontal motion. 

Furthermore, we quantified the fraction of vibrational modes with $p_r$ below the lowest value for a monodisperse sample ($c_v=0$), as shown by the dashed red lines in Fig.\ \ref{Sim}D-E. This fraction increased monotonically with the amount of disorder and the ratio of vertical to horizontal confinement, $k_v/k_h$. As illustrated in Fig.\ \ref{Sim}H, for strong confinement, the gap between the frequency bands is large and a small amount of disorder can quickly lead to mode localization. For weak confinement, the gap is small or nonexistent and mode localization requires more structural disorder. For all the data shown in Fig.\ \ref{Sim}H, intermittent switching occurs when the fraction of low-$p_r$ modes is approximately 0.07--0.2. In this manner, the linear, equilibrium properties can inform system's nonequilibrium dynamical response \cite{Mukamel2000}.

In order to better characterize the dynamical behavior, we identified 5 relevant dimensionless numbers through Buckingham's Pi theorem: $c_v$, $k_v/k_h$, $U_0/k_h\lambda^2$, $k_v /m\gamma^2$, and $\phi/k_v\lambda^2\gamma$. $c_v$ characterizes the degree of structural disorder. $k_v/k_h$ and $U_0/k_h\lambda^2$ determine the degree of quasi-2D confinement. If either number is too small, then the system's equilibrium configuration will be ``buckled" into the $z$-direction (Fig.\ S2 \cite{supp}). $k_v /m\gamma^2$ determines the quality factor of underdamped vertical oscillations. Lastly, $\phi/k_v\lambda^2\gamma$, is the only number associated with the external forcing. This can be intuited as the amount of noisy power necessary for a stochastic, damped  harmonic oscillator to reach an average amplitude $\lambda$, $k_v\lambda^2/2=\phi/4\gamma$ (Fig.\ S6) \cite{supp}. This amplitude threshold will induce rearrangements in the crystalline lattice since the particle spacing is also of order $\lambda$, similar to the Lindemann criteria in classical melting \cite{Lindemann1910}. For a thermal, Brownian oscillator, $2\gamma k_B T$ would assume the role of $\phi$ in our simulations.

\begin{figure*}[t!]
\begin{center}
\includegraphics[width=6.7 in]{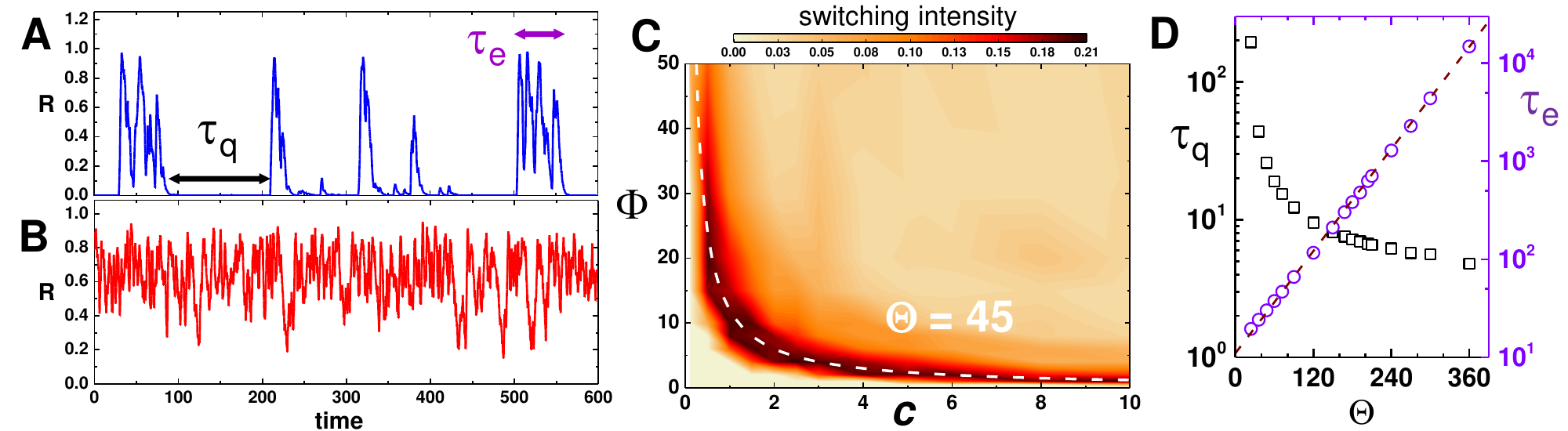}
\caption[]{ Temporal evolution of fractional horizontal mechanical energy, $R$ exhibits intermittent switching (A) for $c=1$ and $\Phi=8$ and perpetually melted state (B) for $c=4$ and $\Phi=20$, $\gamma=0.5$ in both cases. (C) Heatmap showing the switching intensity of R as a function of c and $\Phi$ with $\gamma=0.5$. The dotted line corresponds to a constant $\Theta=45$ and captures the inverse relationship between $\Phi$ and c. (D) The mean duration of quiescent ($\tau_q$) and excited ($\tau_e$) states for varying $\Phi$ (with $\gamma=0.5$ and $c=1.5$). The states are separated using threshold algorithm on the quantity A. The functional form of the dashed line is $0.11\exp(-0.12\Theta)$.
\label{minmod} }
\end{center} 
\end{figure*}

We can form a single dimensionless number that describes the system-wide, intermittent dynamics by considering the role of disorder on the vertical oscillations. Since each particle has a slightly different mass, their vertical frequencies vary, $\omega_v+\Delta\omega_v=\sqrt{k_v/(m+\Delta m})\sim\omega_v(1-c_v/2)$, where $c_v=\Delta m/m$. For $\Delta\omega_v$ to be significant, it must be larger than $\gamma$, which determines the broadness of their response in frequency space. If $\gamma$ is too large, the oscillators will be strongly coupled. 
Thus, by multiplying the stochastic forcing, $\phi/k_v\lambda^2\gamma$, by the disorder in frequency space, $c_v \omega_v/\gamma$, we obtain
\begin{equation}
\theta=\dfrac{c_v\phi}{\sqrt{k_v m}\lambda^2\gamma^2}.
\end{equation}
$\theta$ characterizes the ratio of energy input to dissipation, modulated by disorder. This is analogous to a ``friction factor" in transitional pipe flow that includes the cooperative effects of wall roughness and Reynolds number \cite{Goldenfeld2006}. For reasonable values of the other dimensionless numbers, intermittent dynamics should occur along contours of constant $\theta$, as shown by the white dashed line in Fig.\ \ref{Sim}C (Fig. S5 \cite{supp}).

\textit{Minimal Model for Many-Body Turbulence} --- An ecosystem of two or more competing species is an archetypal model for studying oscillations in dynamical systems \cite{Lotka1910,Volterra1926}. In addition to intermittent turbulence in pipe flow \cite{Shih2016}, other recent examples include the oscillation of the light emission from dust-forming plasmas \cite{Ross2016}, and intermittent precipitation in climate models \cite{Koren2011}. In our many-body system, we can derive a two-species framework by considering the total horizontal mechanical energy ($A$) as a predator, and the total vertical mechanical energy ($B$) as a prey to be consumed:
\begin{align}
\label{Aeq}
\frac{dA}{dt} &=-\gamma A+cAB\left(1-\sqrt{A/B}\right),
\\ \frac{dB}{dt} &=-\gamma B-cAB\left(1-\sqrt{A/B}\right)+w(t)\sqrt{B\Phi/\Delta t}.
\label{Beq}
\end{align}

Here we have assumed that kinetic and potential energies are equipartitioned in each direction, so that $A=2\left<KE_{xy}\right>$ and $B=2\left<KE_{z}\right>$. The first terms on the right hand side in \ref{Aeq} and \ref{Beq} correspond to the power dissipated through hydrodynamic damping, $\sum\vec{\bf F}_d\cdot\vec{\bf v}=-\sum\gamma m (v_x^2+v_y^2+v_z^2)=-\gamma(A+B)$, where the sum runs over each particle. The second term, derived from classical scattering theory \cite{supp}, characterizes ``predation'' and obeys energy conservation. The constant $c$ controls the coupling between vertical and horizontal energies, and parameterizes the polydispersity in the many-body system. Finally, the third term in Eq.\ \ref{Beq} represents the instantaneous power delivered in the vertical direction, $\vec{\bf F}_s\cdot v_z\hat{\bf z}$, where $v_z\sim \sqrt{B/m}$ and $\Phi$ is the average power. Similar noise terms are commonly used to model demographic stochasticity in ecological systems, often resulting in population oscillations \cite{Mckane2005}.

These equations recreate the three observed dynamical regimes. To directly compare with the numerical simulations, we define $R\equiv A/(A+B)$ as the fractional horizontal mechanical energy, in analogy with $\Delta_{xy}$ (Fig.\ \ref{Sim}B).  For intermediate values of $c$ and $\Phi$, $R$ exhibits intermittent behavior (Fig.\ \ref{minmod}A), whereas sufficiently increasing either parameter results in a perpetually excited, equipartitioned state ($A\sim B$) (Fig.\ \ref{minmod}B). The dependence of the switching intensity on both parameters can be seen in Fig.\ \ref{minmod}C. The heatmap shows that there is a distinct region where intermittent dynamics can be observed. In a similar manner to the particle-based simulations, this regime can be described by a single dimensionless number, $\Theta= c \Phi/\gamma^2$, which represents the ratio of the energy input and coupling to dissipation. Intermittent dynamics are only observed for  $20\lesssim \Theta \lesssim120$ (Fig.\ \ref{minmod}C and Fig.\ S8 \cite{supp}), in analogy to transitional pipe flow, where intermittent turbulence is observed for intermediate Reynolds numbers ($1700\lesssim$ Re $\lesssim 2300$) \cite{Kerswell2005}.

Furthermore, transitional pipe flow is known to exhibit spatiotemporal, critical behavior \cite{Goldenfeld2017}. The lifetime statistics and universality class are related to directed percolation where laminar flow acts as a nonequilibrium absorbing state \cite{Shih2016,Goldenfeld2017,Pomeau1986,Pomeau2016,Barkley2016}. For intermittent turbulence with spatially-separated, coexisting, independent ``puffs", the lifetime of the turbulent state scales super-exponentially with Reynolds number \cite{Avila2011,Shih2016}. For a given $\Theta$ in our minimal model, the probability distribution of both excited ($\tau_e$) and quiescent ($\tau_q$) state lifetimes have exponential tails (Fig.\ S9 \cite{supp}), and the mean lifetime of the excited state scales exponentially with $\Theta$ (Fig.\ \ref{minmod}D). This scaling originates from the memoryless noise driving the system. If our many-body system were spatially-extended and could support multiple excited, gas-like regions, we would expect extreme value statistics and subsequently, super-exponential behavior \cite{Goldenfeld2017}.

\textit{Summary} --- Intermittent dynamics are commonly observed in natural systems but rarely understood. Here we presented a distinct mechanism to promote intermittent dynamics in a tractable, many-body system. A layer of polydisperse, interacting particles, damped by the environment and driven anisotropically by noise, can intermittently transition between gaseous and crystalline states. Spatial heterogeneities, in the form of localized vibrational modes, couple with the external noise to facilitate this transition. In this sense external noise acts as an engine that drives the system in one direction, and structural heterogeneities act as a rudder that intermittently steers energy into other degrees of freedom.  We also derived a single dimensionless number that characterizes the intermittent regime, in analogy to the Reynolds number in transitional pipe flow. Moreover, we provided a minimal model that captures the essential features of the many-body dynamics, and demonstrates a exponential scaling of the lifetimes of the excited states, a feature only recently confirmed for the lifetime of turbulent ``puffs" in fluid flow. We hope these results lead to further connections between simple, discrete particle dynamics and natural complex systems.

\textit{Acknowledgements} --- We acknowledge financial support from the National Science Foundation Grant No. 1455086. We gratefully acknowledge H.-Y. Shih and N. Goldenfeld for stimulating discussions about predator-prey systems. We would like to thank J. Mendez, M. Kawamura, H. Saul, J. Silverberg, A. Roman, I. Nemenman, M. Martini, B. Beal, and B. Doolittle for useful discussions.

\bibliographystyle{apsrev4-1}

\end{document}